\documentclass[lettersize,journal]{IEEEtran}

\usepackage{amsmath,amsfonts}
\usepackage{algorithmic}
\usepackage{algorithm}
\usepackage{array}
\usepackage[caption=false,font=normalsize,labelfont=sf,textfont=sf]{subfig}
\usepackage{textcomp}
\usepackage{stfloats}
\usepackage{hyperref}
\usepackage{url}
\usepackage{verbatim}
\usepackage{graphicx}
\usepackage{cite}
\begin{document}

\title{EMC Limit Level Guidelines for In-System Interference with GPS Receivers}

% DB: please add authors

\author{Giorgi~Tsintsadze,~\IEEEmembership{Member,~IEEE,}
        Haran~Manoharan,~\IEEEmembership{Student Member, IEEE},
        Aaron~Harmon,~\IEEEmembership{Student Member, IEEE},
        Daniel~Commerou,~\IEEEmembership{Member, IEEE},
        Connor~Buneta,~\IEEEmembership{Student Member, IEEE},
        Brian~Booth,~\IEEEmembership{Member, IEEE},
        Daryl~Beetner,~\IEEEmembership{Fellow,~IEEE}% <-this % stops a space
        
\thanks{Manuscript received XXX; This work was supported in part by the National Science Foundation (NSF) under Grant IIP-1916535.}% <-this % stops a space
% DB: Need to fill in author information

\thanks{G. Tsintsadze, Haran Manoharan, Aaron Harmon, Daniel Commerou, Connor Buneta and D. Beetner are with the EMC Laboratory, Missouri University of Science and Technology, Rolla, MO 65409 USA (e-mail:  daryl@mst.edu).}
\thanks{B. Booth is with John Deere Electronic Systems, Fargo, ND USA.}}



\maketitle

\begin{abstract}
Because GPS signals are weak, electronic systems and components that are placed near GPS receivers can easily cause disruptive electromagnetic interference through their unintended radiated emissions. In this paper, EMC limit level guidelines are presented for electronics that are intended to be placed near to GPS receivers, as often happens in automotive and other applications. One of the challenges of defining limit-levels for systems intended to be integrated with GPS receivers is that the impact of noise at the input of the receiver may vary substantially depending on the form of the noise due to the correlator function implemented by GPS receiver. The quality of the correlated signal is typically represented using the carrier-to-noise ratio ($C / N_0$). A theoretical model predicting the degredation of the carrier-to-noise ratio with radio frequency interference is presented in this paper and is validated with realistic noise sources.  The model is then used to develop guidelines to assess the impact of unintended emissions from electronic devices on nearby GPS receivers based on the frequency, bandwidth, and magnitude of the noise. These guidelines provide a more nuanced method of evaluating emissions than simple limit lines that are used by many emissions standards.   
\end{abstract}


\section{Introduction}
\IEEEPARstart{B}{ecause}
 global positioning system (GPS) signals are weak (e.g. -130 dBm), radio-frequency interference (RFI) can be highly disruptive to receiver operation. RFI can affect RF front-end blocks such as the Automatic Gain Control (AGC) and carrier code tracking loops \cite{ward}, and at higher levels can cause the receiver to completely drop satellite connections.  This interference is particularly problematic when electronic components are integrated near to the GPS receiver, as their unintended emissions, while low, might still be able to disrupt the normal operation of the receiver. Determining an acceptable level of in-system interference is often difficult, as GNSS receivers use a cross-correlation block as their RF front-end and the correlator output can vary significantly for different noise types \cite{conference_paper}. 
 A broad-band signal and a narrowband signal at the same level, for example, may have significantly different effects on GPS performance. While emission standards often set a predefined threshold for the level of emissions allowed in the GPS bands, those thresholds do not consider the types of possible interference, and so are necessarily either too conservative or too lenient, depending on the type of noise. 
 %DB: reference? GT: I think only reference is this paper itself 
 The purpose of this paper is to lay the foundation for a more intelligent evaluation of acceptable levels of in-system RFI when working with GNSS receivers. 

Mathematical expressions for the output power of the correlator for different types of interference signals were presented in \cite{bek} and in \cite{balalei} for continuous wave  (CWI) and pulsed interference. While these expressions are very useful, the formulations were either only validated through simulations or with limited measurement data, and no extensions were made to estimate acceptable levels of in-system interference. In \cite{john_deere} the authors provide  comprehensive measurement data for the carrier-to-noise ratio, $C/N_0$ \cite{ward}, and AGC values under different types of interference, as created by a signal generator. These results give an intuitive understanding of the impact of interference, but do not back up the results with theory. The following paper will validate the theory presented in \cite{conference_paper} with a variety of real-world noise sources and will extend the theoretical results to introduce guidelines for evaluating the impact of noise from electronics placed near to a GPS receiver, depending on the spectral distribution of that noise.

\section{Theory} \label{Theory}

\begin{figure}
  \centering
  \includegraphics[width=0.5\textwidth]{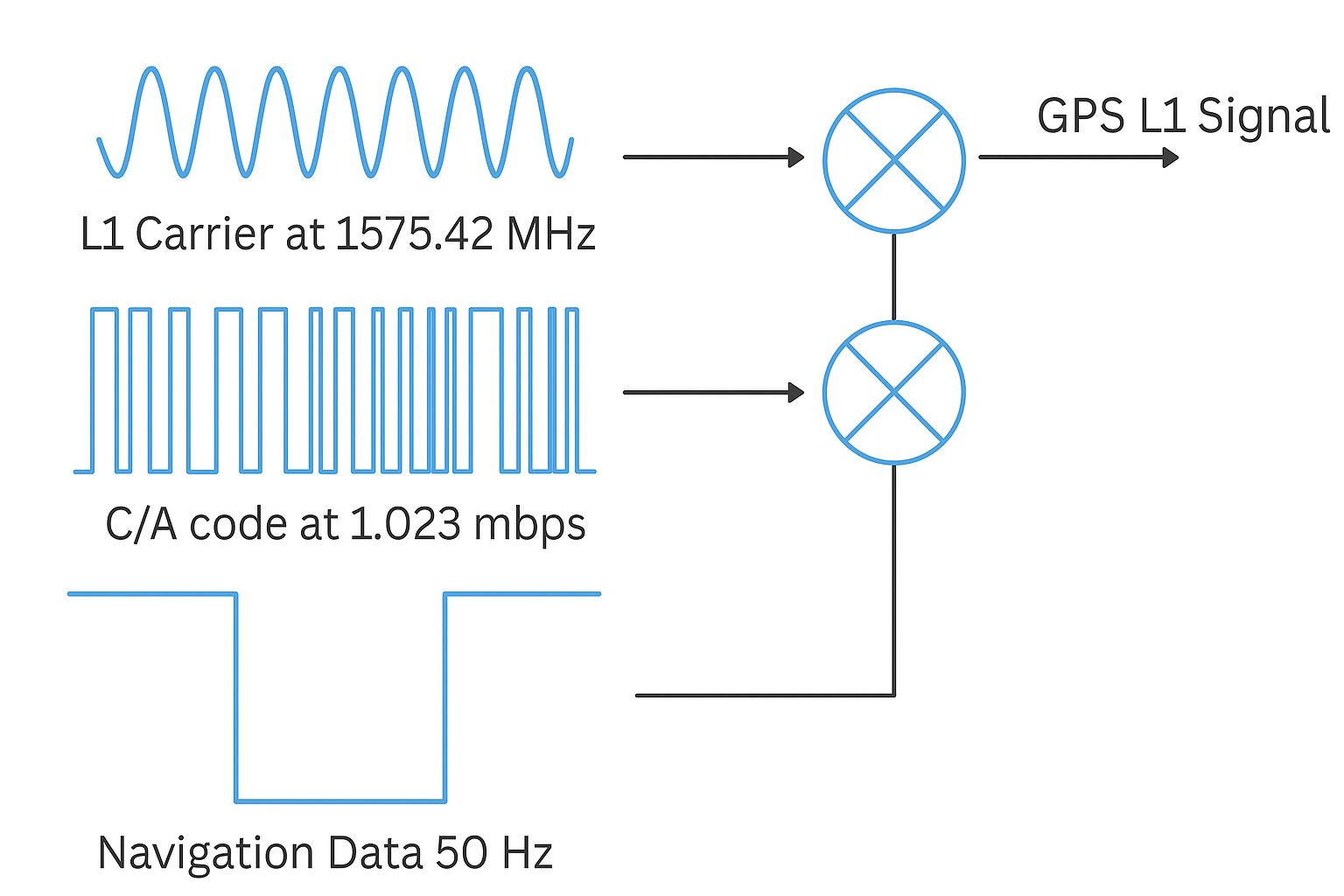}
  \caption{Block diagram of GPS signal modulation}
  \label{fig:gps_signal_architecture}
\end{figure}

GPS satellites use a code division multiple access (CMDA) technique, which means that the navigation data is modulated not only with a carrier signal but also with pseudo-random noise codes (PRN), called C/A codes for GPS L1 signals \cite{ward}. Fig. \ref{fig:gps_signal_architecture} shows a block diagram for the modulation of GPS L1 signals. The C/A codes are called Gold codes because they have an auto-correlation peak of unity while having an ideal cross-correlation and misalignment auto-correlation of zero. This property means that satellites can be identified by their unique PRN and that multiple satellite signals can be received simultaneously even though their carrier frequencies are the same. The long correlation sequence also means that relatively weak signals can be picked out of significant noise - in many cases even well below the noise floor of the receiver. Since C/A codes are finite (consisting of 1023 chips), the cross-correlation is not exactly zero between codes, however the maximum misalignment auto-correlation result or cross-correlation result is 6.35\% of the value found with auto-correlation of the C/A code when properly aligned \cite{bek}.

\begin{figure}
  \centering
  \includegraphics[width=0.5\textwidth]{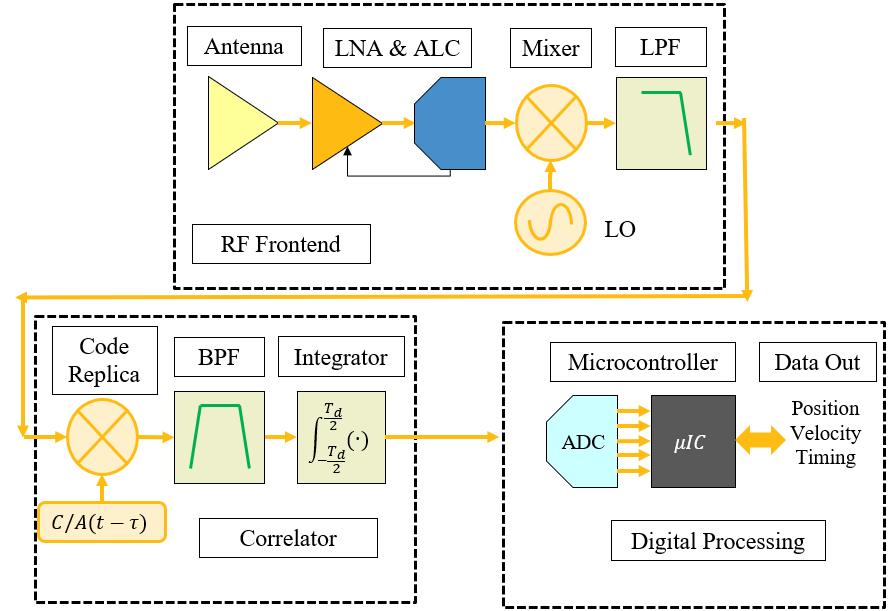}
  \caption{Block diagram of the front-end and correlator.}
  \label{fig:correlator}
\end{figure}

C/A ``bits" are referred to as chips since they do not actually carry information on their own. The C/A chip rate for GPS L1 signals is 1.023\ MHz \cite{bao}. The receiver stores replicas of the PRNs (i.e. C/A codes) transmitted by the satellites and uses them in the correlator block. Fig. \ref{fig:correlator} shows a block diagram of the correlator. The GPS signals first enter the RF front-end, which consists of an antenna, LNA and AGC modules. The front-end is followed by a correlator block.  

\IEEEpubidadjcol

In this paper, the post-correlation signals will be used to estimate the values of $C/N_0$. The following analysis show the output power of the correlator when evaluating a GPS signal in the presence of interference noise. The signal at the RF front-end can be
represented as \cite{bek}:
% DB: Should we also cite Balaei?
%
\begin{align*}
    R_{in}(t) &= S_{\text{PRN}}(t) + I(t) + N_0(t)    
\end{align*}
where $S_{\text{PRN}}(t)$ is the GPS signal (here we will assume a GPS L1 signal), $I(t)$ is the interference signal, and $N_0(t)$ is zero-mean thermal noise. The correlator output will be evaluated below for both narrowband continuous wave interference (CWI) and for (continuous) multitone interference.

\begin{figure}
  \centering
  \includegraphics[width=\linewidth]{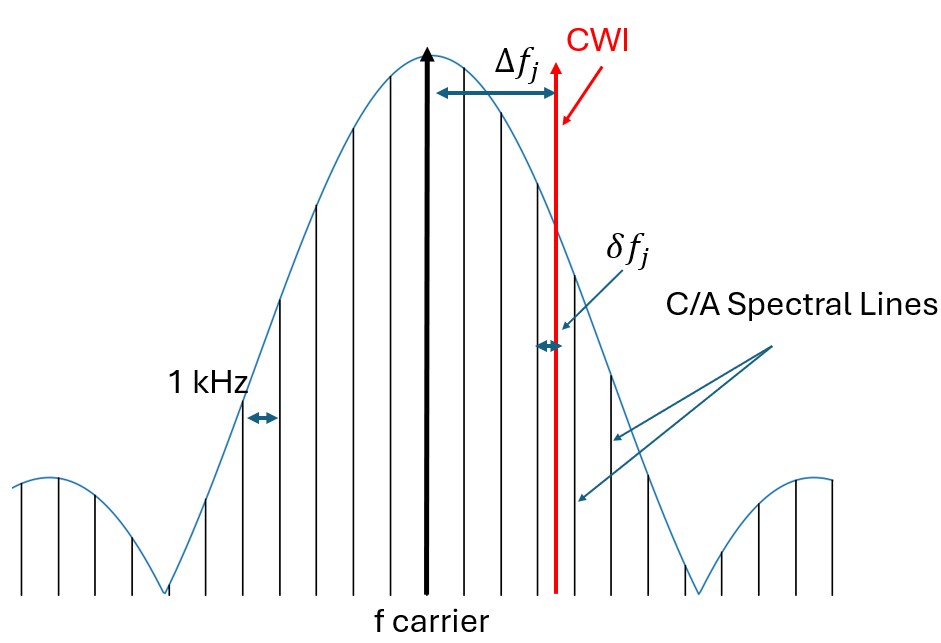}
  \caption{Spectral lines associated with the C/A code and with the continuous wave interference (CWI). $\delta f_j$ is the residual frequency between the closest C/A-code spectral line and the CWI while $\Delta f_j$ is the difference between the CWI signal frequency and carrier or LO frequency.}
  \label{fig:spectral_lines}
\end{figure}

CWI in the instantaneous form can be represented as:
\begin{equation}\label{eq:cwi_signal}
    I(t) = \sqrt{2P_j} e^{j2 \pi f_j t + \theta}
 \end{equation} 
where $P_j$ is the interference power and $f_j$ and $\theta$ are the interference frequency and angle, respectively. A detailed derivation and validation of the output of the correlator block is shown in \cite{conference_paper}, as derived from \cite{bek}, when the noise signal is a single tone sinusoidal continuous wave. In this case, correlator output power is given by:
\begin{equation}\label{eq:cwi_corellator_output}
  P_{\text{CWI}} = |\sqrt{ 2 P_j}\, C_k^* \mathrm{sinc}({ { \pi\,  \delta f_j\, T_d)}} e^{j\theta_{r}}|^2
 \end{equation} 
where, $\delta f_j$ is the difference between the frequency of the interfering signal and the closest spectral line frequency as shown in Fig. \ref{fig:spectral_lines} (spectral lines are spaced by 1 KHz tones), while $ C_k$ are the Fourier series coefficients of the C/A code and are normalized such that 
\begin{equation}\label{eq:coefficients_power_unity}
\sum_{i = 0}^{\infty} |{C_i}|^{2} = 1
\end{equation}. 
 The term {$e^{j\theta_{r}}$ is a simplified form of $ e^{j \theta}  e^{j { 2\pi \frac{i}{T_s} \tau  }}$, since the output of the correlator is taken when the value of $\tau$ results in the alignment of the replica with the GPS signal. Having a complex exponential in the absolute value operator as in (\ref{eq:cwi_corellator_output}) seems redundant as it evaluates to unity, however, as we shall see, when taking multitone signals into consideration, all individual interference must be first summed and can then be subsequently squared to calculate the correlator power output; in that case, phase differences cannot be ignored.   
The integration time, $T_d$, is a receiver design parameter, and in this particular case is a known value. For uncorrelated zero-mean noise, a longer integration time yields a higher alignment peak output, and hence a higher $C/N_0$. In-band CWI is not necessarily uncorrelated noise, but  (\ref{eq:cwi_corellator_output}) shows that increasing the integration time reduces the noise power induced by the CWI when there is a difference of $\delta f_j$ between the spectral lines of the C/A code and the CWI frequency. In rare cases, where the CWI frequency is an exact integer multiple of 1~KHz away from the carrier and sits on top of one of the spectral lines, increasing the integration time will not reduce the noise, since there is a true correlation between the C/A code and the CWI. While this precise overlap is rare for truly narrowband CWI, it should be noted that even a 1 kHz bandwidth of the interfering signal will guarantee overlap between the interfering source and the C/A code spectra.

Equation (\ref{eq:cwi_corellator_output}) can be extended for noise with a general spectrum by recognizing two things. First, that the integrator is acquiring signals for time $T_d$, and therefore any continuous spectrum noise will become noise with discrete spectral lines with spacing of $\frac{1}{T_d}$ from the integrator's perspective. Second, that the correlator block is a linear system and therefore it is possible to evaluate the correlator output in the frequency domain for each noise spectral line and then sum the results to obtain the correlator output power for the given noise. Using this approach, the signal-to-noise ratio in the correlator output can be calculated as:
\begin{equation}\label{eq:cn_multitone}
   \begin{split}
   \text{S}&\text{NR(dB)} = \\
    &10\ \mathrm{log}({ \frac{P_s R(\tau)\, \mathrm{sinc}(\pi\, \Delta f\, T_d)^2}{ N_0 + | \sum_{n=1}^{N}\sqrt{2 P_n} C_n^*\, \mathrm{sinc}({ { \pi\,  \delta f_j\, T_d)}} e^{j\theta_{r}}|^2}}) 
   \end{split}
\end{equation} 
where N is number of interference tones, $R(\tau)$ is an autocorrelation function between the incoming signal and the local code replica and is unity when $\tau$ is such that replica aligns with incoming signal, $\Delta f$ is the Doppler shift between the satellite and the receiver local oscillator. Modern ASICs have a multidimensional search space where they not only find the correct value for $\tau$ to align the local code replica with the signal, but are also sweeping frequency to maximize signal power with Doppler shift. As a result, $R(\tau) \mathrm{sinc}(\pi \Delta f T_d)$} can be regarded as unity.
The carrier-to-noise ratio can be calculated from the SNR as:
\begin{equation}\label{eq:cn_multitone_2}
    C/N_0 (dB/Hz) = \text{SNR}(dB) + 10\ \mathrm{log}(\frac{1}{T_s})
\end{equation} 
Notice that $C/N_0$ is the signal-to-noise ratio evaluated over a $1/T_s$ bandwidth.

\subsection{Sinc envelope approximation}

\begin{figure}
  \centering
  \includegraphics[width=0.52\textwidth]{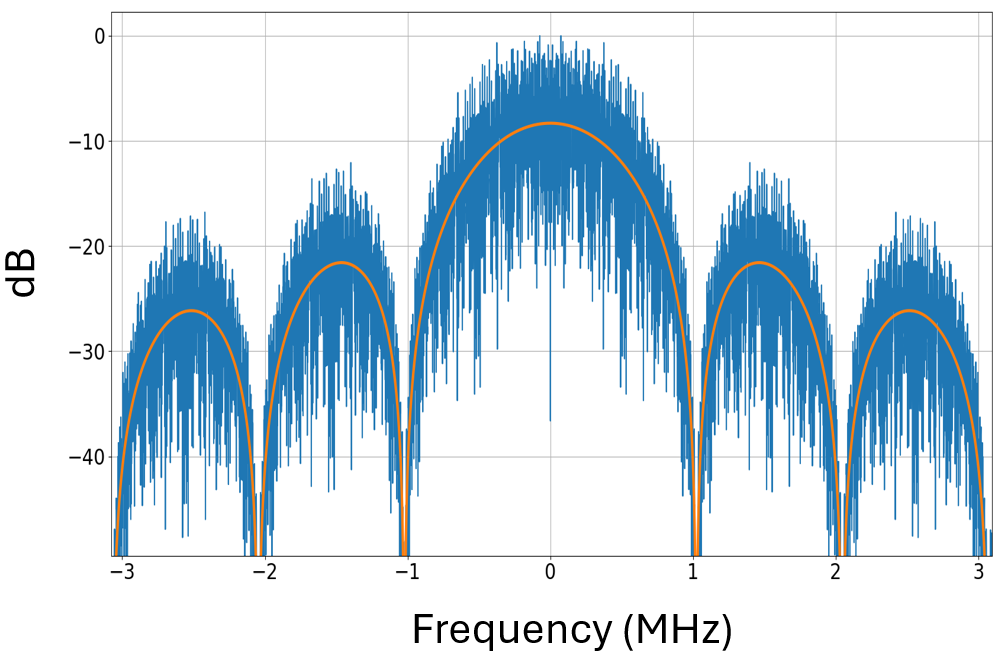}
  \caption{Sinc envelope approximation of the C/A Code. The plot is normalized to the magnitude of the largest spectral line.}
  \label{fig:sinc lobe}
  \label{fig:sinc}
\end{figure}

While (\ref{eq:cwi_corellator_output}) provides a useful expression for evaluating the power output of the correlator for a specific CWI signal and a specific C/A code, it is challenging to generalize it for interference from all satellites in its current form, as  $C_{w}$ corresponds to the spectra from a particular code. To suggest interference guidelines for a variety of satellites, it is reasonable to assume the noise bandwidth is larger than 1 kHz (so that it will always overlap at least one C/A code spectral line) and to consider the average values of C/A spectral lines over all codes. In this case, the C/A code spectra can be approximated with a sinc envelope as shown in Fig. \ref{fig:sinc} with the form:
\begin{equation}\label{eq:sinc_envelope}
     A_{0}\frac{sin(\pi f T_{a})}{(\pi f T_{a})}
\end{equation} 
where $T_{a}$ is the duration of the chip sequence (1 us for GPS) and $A_{0}$ is chosen such that:
\begin{equation}\label{eq:cn_multitone_2}
    \int_{- \infty}^{\infty} A_{0}^2\frac{sin^2(\pi f T_{a})}{(\pi f T_{a})} df = 1, 
\end{equation}
meaning that the total power corresponding to the sinc envelope is the same as the power associated with the Fourier-series coefficients (\ref{eq:coefficients_power_unity}). 
% DB: I believe you're missing an equation here? This reference cannot be found.
Using this sinc envelope approximation, (\ref{eq:cwi_corellator_output}) can be rewritten as:
\begin{equation}\label{eq:cwi_output_sinc}
P_{\text{CWI}} \approx  |\sqrt{2 P_j} A_{0} \cdot sinc(\pi f_i T_{a}) \cdot sinc(\pi\, \delta f_j\, T_d) e^{j \theta_r}|^2
\end{equation} 
where $f_i$ is the interference frequency, esentially evaluating sinc value at interference frequency and approximating it as a value for a spectral line, while the latter depends on exact C/A sequence, in this way we are approximating expected value at that frequency. 
% DB: is the above statement correct?
Equation (\ref{eq:cwi_output_sinc}) can then be used in (\ref{eq:cn_multitone}) instead of (\ref{eq:cwi_corellator_output}) to calculate $C/N_0$ from a multitone signal using the sinc approximation of C/A spectral lines. Both theories will be compared to measurement in Section IV. 

\section{Measurement Setup} \label{Measurement}

\begin{figure}
  \centering
  \includegraphics[width=0.5\textwidth]{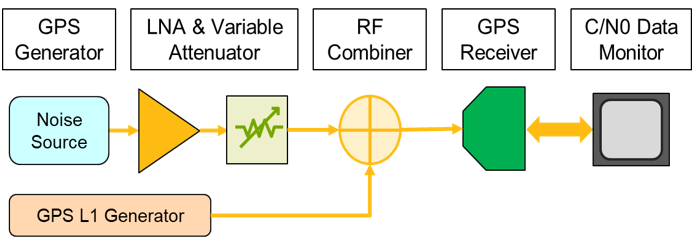}
  \caption{Block diagram of measurement setup. SDR stands for Software Defined Radio and AWG stands for Arbitrary Waveform Generator.}
  \label{fig:measurement_blockdiagram}
\end{figure}

The equations derived above were experimentally validated by applying a GPS L1 signal combined with a noise source to a commercially available GPS receiver. Fig. \ref{fig:measurement_blockdiagram} shows a block diagram of the measurement setup. The GPS receiver allowed $C/N_0$ and some other signal parameters to be accessed using special software provided by the manufacturer. To ensure the precision of the measurements, $C/N_0$ was only evaluated after the GPS receiver was successfully locked to the incoming satellite signals. To allow the receiver to reach a "locked" state, a realistic GPS signal must be used as an input to the receiver.  

\subsection{GPS Signal and Noise Generation}

A replica of a typical constellation of GPS satellite signals was generated using a BladeRF software defined radio (SDR) running open-source software available on Github made for this purpose \cite{github}.
% DB: we are missing this citation
The GPS simulation file was generated using a sample rate of 10\ MSPS instead of default values of 2\ MSPS and setting a flag in the software to maintain constant satellite power, without accounting for additional free space loss due to a satellite's movement over horizon.  The "multi-path" option on the receiver was also disabled. In \cite{conference_paper} theory was validated through measurements where multi-tone interference signal was generated through arbitrary waveform generator and phase for each tone is set to random values and averaged over 1000 sample instantiations to get averaged for each C/N value for each power level. In this paper though, the theory is extended and also validated using "real-life" interference signals conceived from HDMI cable and switching power supply. On top of that theory is extended to establish baseline acceptable interference levels and estimations for acceptable interference levels for different types of noises.

\section{Results with Real-Life Noise Sources}

\begin{figure}
  \centering
  \includegraphics[width=0.4\textwidth]{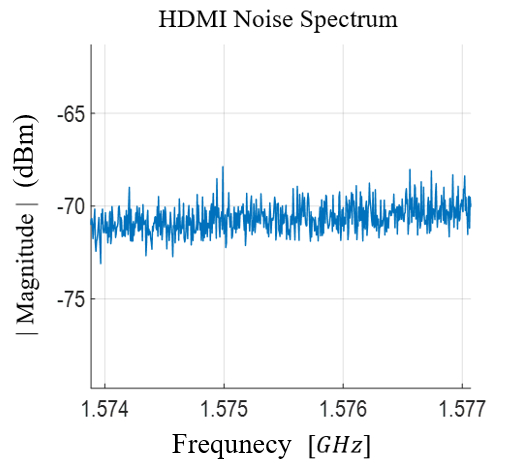}
  \caption{HDMI cable noise level without attenuation.}
  \label{fig:hdmi_noise}
\end{figure}

\begin{figure}
  \centering
  \includegraphics[width=0.5\textwidth]{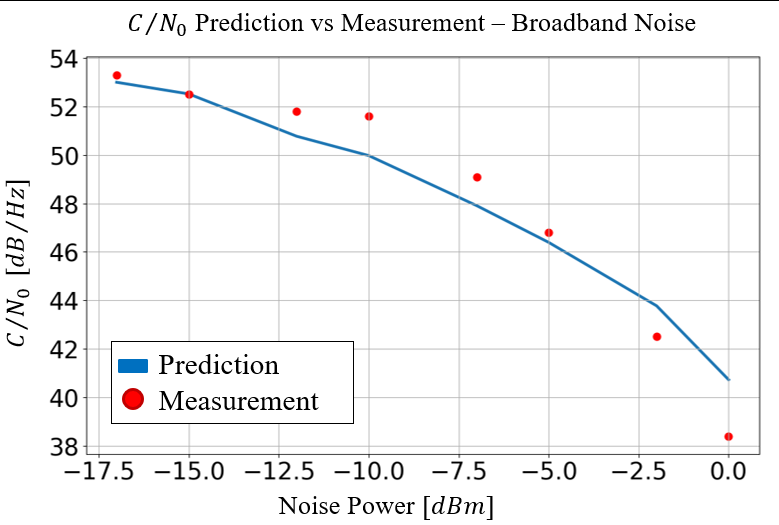}
  \caption{HDMI cable noise measurement and simulation comparison }
  \label{fig:hdmi_measurements}
\end{figure}

Two different noise sources were used to validate the theory against realistic noise sources: a switching power supply and an HDMI cable. Figure \ref{fig:hdmi_noise} shows the power spectral density measured by a current clamp around  the HDMI cable. This source creates a a relatively constant level of broad-band noise. To predict the impact of this noise, the noise measurement was exported from the Spectrum Analyzer and each point was regarded as single-tone interference. Equation (\ref{eq:cn_multitone}) was then used to calculate $C/N_0$. Figure \ref{fig:hdmi_measurements} shows a comparison of the predicted and measured values of $C/N_0$. To allow measurement of $C/N_0$ at multiple noise levels, the measured noise was routed through a variable attenuator before adding it to the GPS signal. The reading "0 dBm" means that no attenuation was applied to the noise read by the current clamp. 

\begin{figure}
  \centering
  \includegraphics[width=\linewidth]{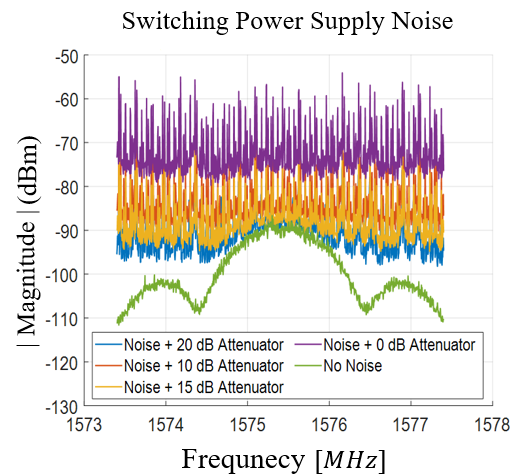}
  \caption{Noise spectrum of power suplly without attenuators}
  \label{fig:power_supply_noise}
\end{figure}

\begin{figure}
  \centering
  \includegraphics[width=\linewidth]{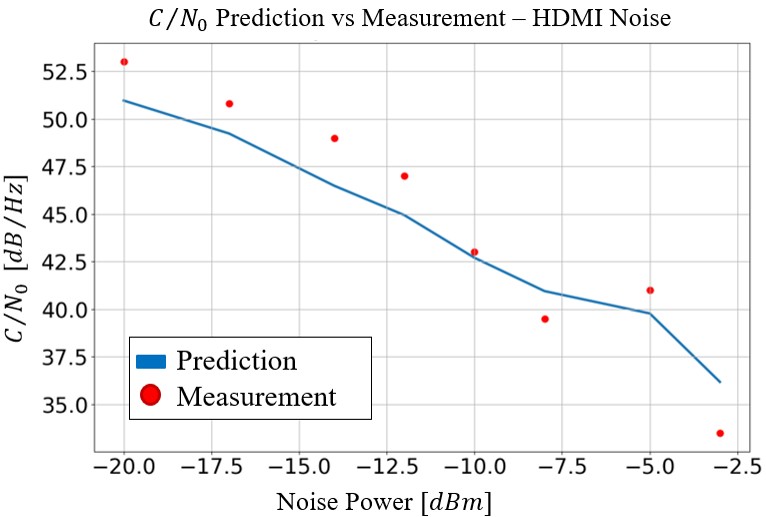}
  \caption{Power supply noise measurement and simulation comparison }
  \label{fig:power_supply_result}
\end{figure}

Fig. \ref{fig:power_supply_noise} shows the noise spectrum from the switching power supply. As with the HDMI case, a variable attenuator was used to vary the power of the noise and measure the degradation of $C/N_0$ with noise power. Fig. (\ref{fig:power_supply_result}) shows the predicted and measured values of $C/N_0$. Overall, for the two noise sources, the model was able to estimate $C/N_0$ within ~2.1\ dB compared to the values found in measurements.

\section{Mesoband Noise}

\begin{figure}
  \centering
  \includegraphics[width=\linewidth]{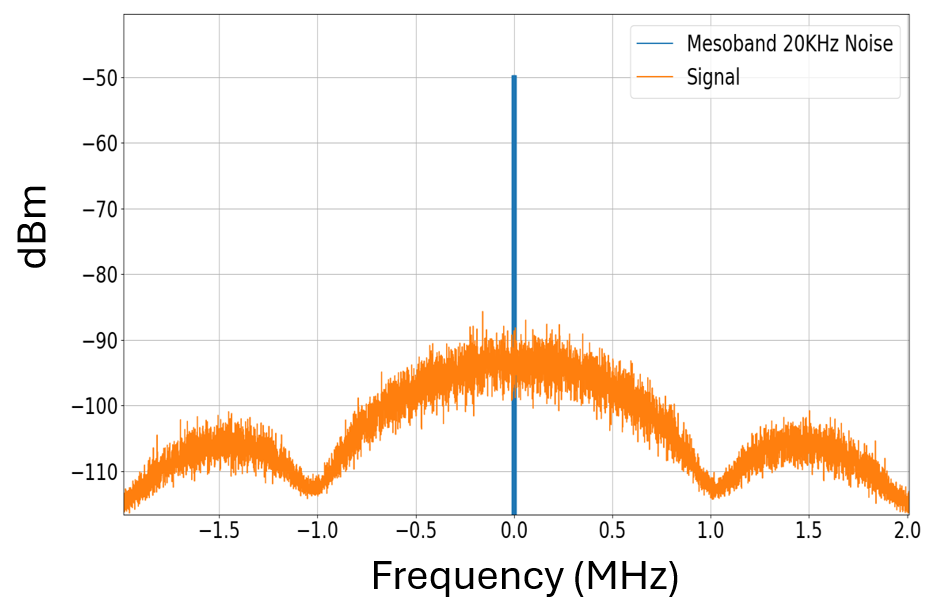}
  \caption{20\ kHz mesoband noise superimposed over the C/A code spectra for GPS L1.}
  \label{fig:mesoband_spectrum}
\end{figure}

For GPS receivers, mesoband noise can be defined as noise with a relatively small bandwidth of more than 1\ kHz but substantially less bandwidth than that of the main lobe of the sinc function for the C/A code (e.g. 1/10). In this case, the noise can be assumed to be constant over this bandwidth and the interaction of the noise with the C/A code spectral line (Fourier coefficient on n-th spectral line where n), $C_n$, can be approximated by just evaluating (\ref{eq:sinc_envelope}) at the center frequency of the mesoband noise and assuming that  other C/A code values are the same over the noise bandwidth. Fig \ref{fig:mesoband_spectrum} shows an example of 20\ kHz mesoband signal at the center frequency of the GPS signal. Since the integration time is finite, the resolution bandwidth of the receiver is also finite and is the reciprocal of the integration time. The integrator output resulting from multi-tone mesoband noise can be approximated as:
%
% DB: is the following equation correct?
\begin{equation}\label{eq:mesoband_integrator_output}
     W_{correlator} \approx \frac{ \sqrt{2 P_0}}{2}{} A_{0}\, sinc(\pi f_{c} T_{a}) \sum_{n = 1}^{N} sinc(\pi\, \delta f_w\, T_{d}) e^{j \theta_w}.
\end{equation} 
Here $A_{0}sinc(\pi f_{c} T_{a})$ represents the approximation of the C/A code specta with a sinc envelope which is evaluated at the center-frequency of the noise and is approximated as constant over the bandwidth of the noise. N stands for number of tones and for continuous band spectrum equals to $BW_{noise} * T_d$. Moreover, $\delta f_w $ can be represented as a random variable that takes values from 0 to 500\ Hz with uniform probability. This is a reasonable approximation as due to doppler shift C/A lines are being shifted while noise keeps the same frequency, resulting in change for $\delta f_w $. 

\begin{figure}
  \centering
  \includegraphics[width=\linewidth]{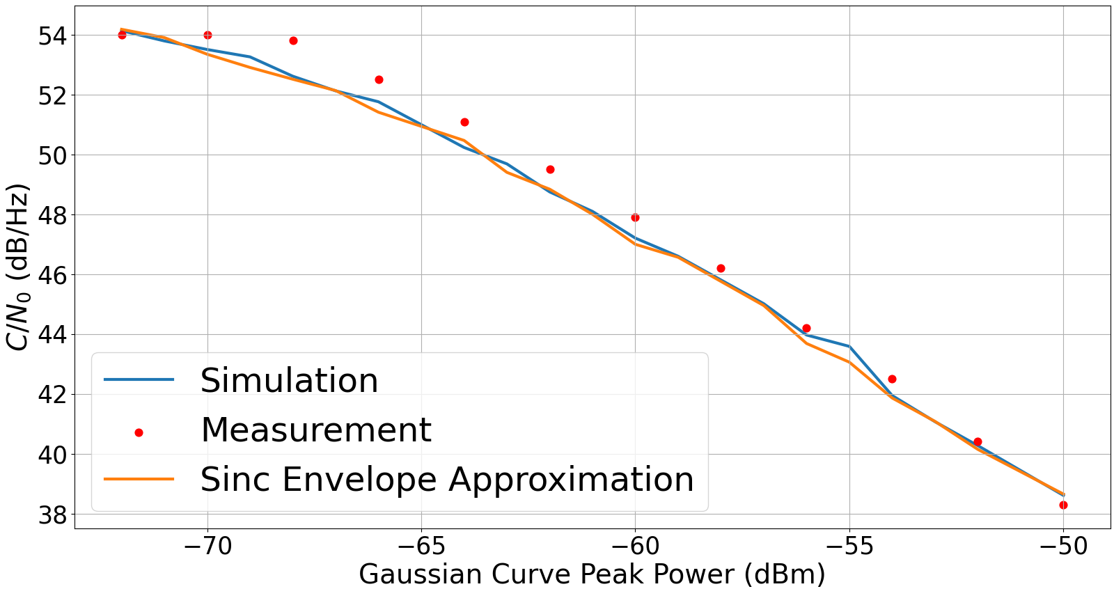}
  \caption{AWG settings 20\ kHz span 42 tones, simulation $T_d$=5ms}
  \label{fig:mesoband_meas_sim}
\end{figure}

Fig. \ref{fig:mesoband_meas_sim} shows the measured and predicted values of $C/N_0$ using (\ref{eq:mesoband_integrator_output}) when  20\ kHz mesoband noise was added at the center frequency of the GPS L1 signal Fig. \ref{fig:mesoband_spectrum}. It should be noted that when generating the mesoband noise, number of tones were set to 42, this was partly due to arbitrary waveform generator instrument limitation combined with the fact that frequency between noise bins would be around 470 Hz which is more than twice as small than 1 KHz of C/A spectral line spacing, guaranteeing at least 1 interference CWI per spectral line.

\begin{figure}
  \centering
  \includegraphics[width=\linewidth]{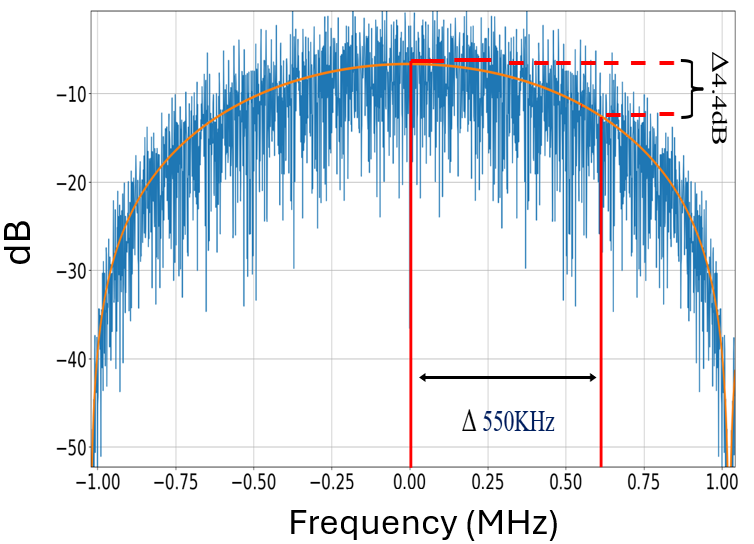}
  \caption{Changing relative difference between the noise center frequency and the GPS carrier frequency from 0\ Hz to 550\ KHz, should decrease impact of the mesoband noise by about 4.4\ dB, as predicted by the sinc envelope.}
  \label{fig:mesoband_550khz_graph}
\end{figure}

\begin{figure}
  \centering
  \includegraphics[width=\linewidth]{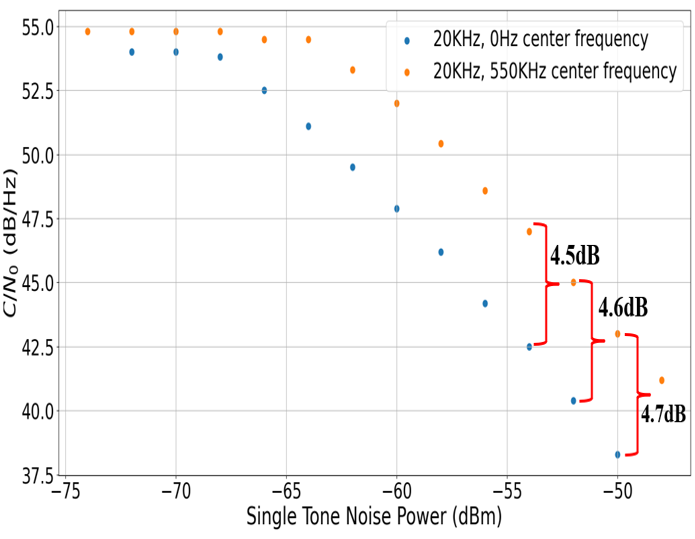}
  \caption{Measured values of $C/N_0$ when 20\ kHz bandwidth noise is centered at the GPS carrier frequency and when it is centered 550\ kHz away.  $C/N_0$ is around ~4.5\ dB higher when the noise is centered at 550KHz, as expected.}
  \label{fig:550Khz_comparison_cn}
\end{figure}

Equation (\ref{eq:mesoband_integrator_output}) makes it easy to mathematically understand the implications of changing the center frequency of the mesoband noise. The effect on $C/N_0$ degradation can be evaluated simply by following the sinc envelope. Fig \ref{fig:mesoband_550khz_graph} shows that moving the noise center frequency from 0\ Hz to 550\ kHz with respect to the carrier frequency results in a change of around -4.4 dB in the value of the C/A code sinc envelope. If the bandwidth and power of the mesoband noise are kept the same but its center frequency changes, it is thus expected that the multiplication of the C/A code and the noise will decrease by approximately the same amount as the change in the value of the sinc envelope, and thus that $C/N_0$ will increase by the same value. Fig \ref{fig:550Khz_comparison_cn} show measured values of $C/N_0$ when 20\ kHz bandwidth noise was added at the GPS code carrier frequency and 550\ kHz from the carrier frequency. The difference in the $C/N_0$ values are roughly 4.4\ dB as expected. 

\begin{figure}
  \centering
  \includegraphics[width=\linewidth]{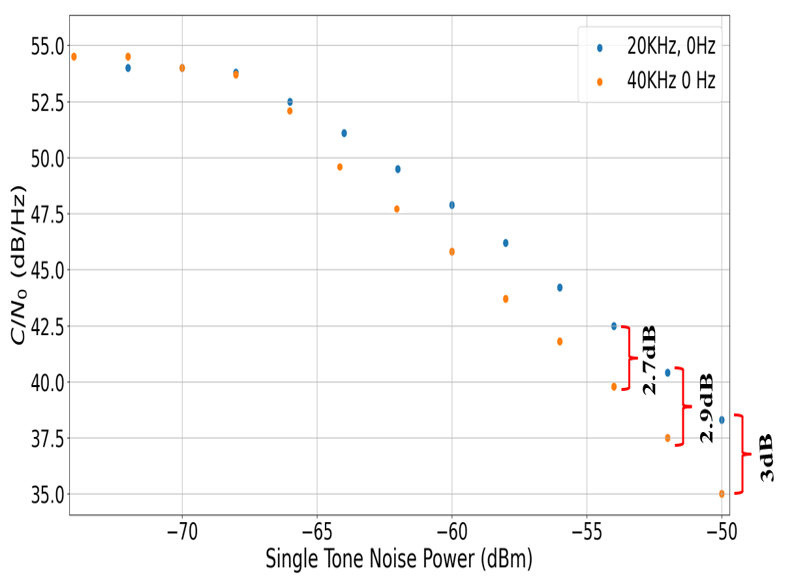}
  \caption{Measured values of $C/N_0$ when mesoband noise at the center frequency of the C/A code is added with 20\ KHz and 40\ kHz bandwidth. $C/N_0$ is roughly 3\ dB lower with 40\ kHz noise than 20\ kHz noise, as expected. }
  \label{fig:40khz_mesoband_results}
\end{figure}

Equation (\ref{eq:mesoband_integrator_output}) also indicates the impact of the mesoband noise bandwidth. Doubling the mesoband bandwidth, for example, should also double the impact of the noise, leading to expected ~3\ dB reduction in $C/N_0$. Fig  \ref{fig:40khz_mesoband_results} shows the measured $C/N_0$ when 40\ kHz and 20\ kHz mesoband noise centered at 0\ Hz from the carrier was added to the signal, resulting in an approximately 3\ dB drop in $C/N_0$ as expected.

\begin{figure}
  \centering
  \includegraphics[width=\linewidth]{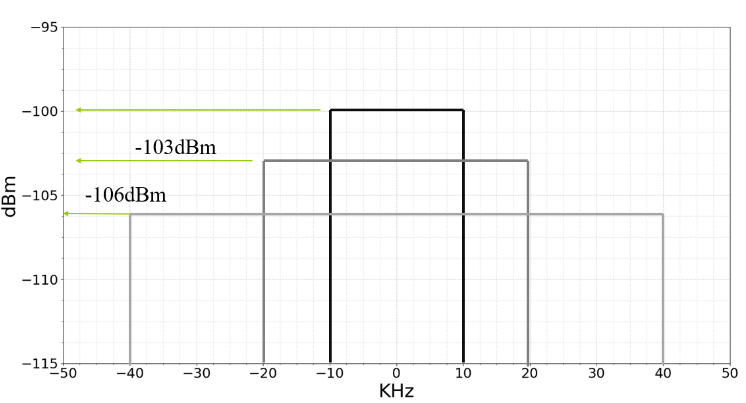}
  \caption{Limil level for rectangular mesoband noise center at GPS L1 frequency for different bandwidths of 20 KHz, 40 KHz and 80 KHz}
  \label{fig:lmit_level_mesoband_const_center_freq_fig}
\end{figure}

\begin{figure}
  \centering
  \includegraphics[width=\linewidth]{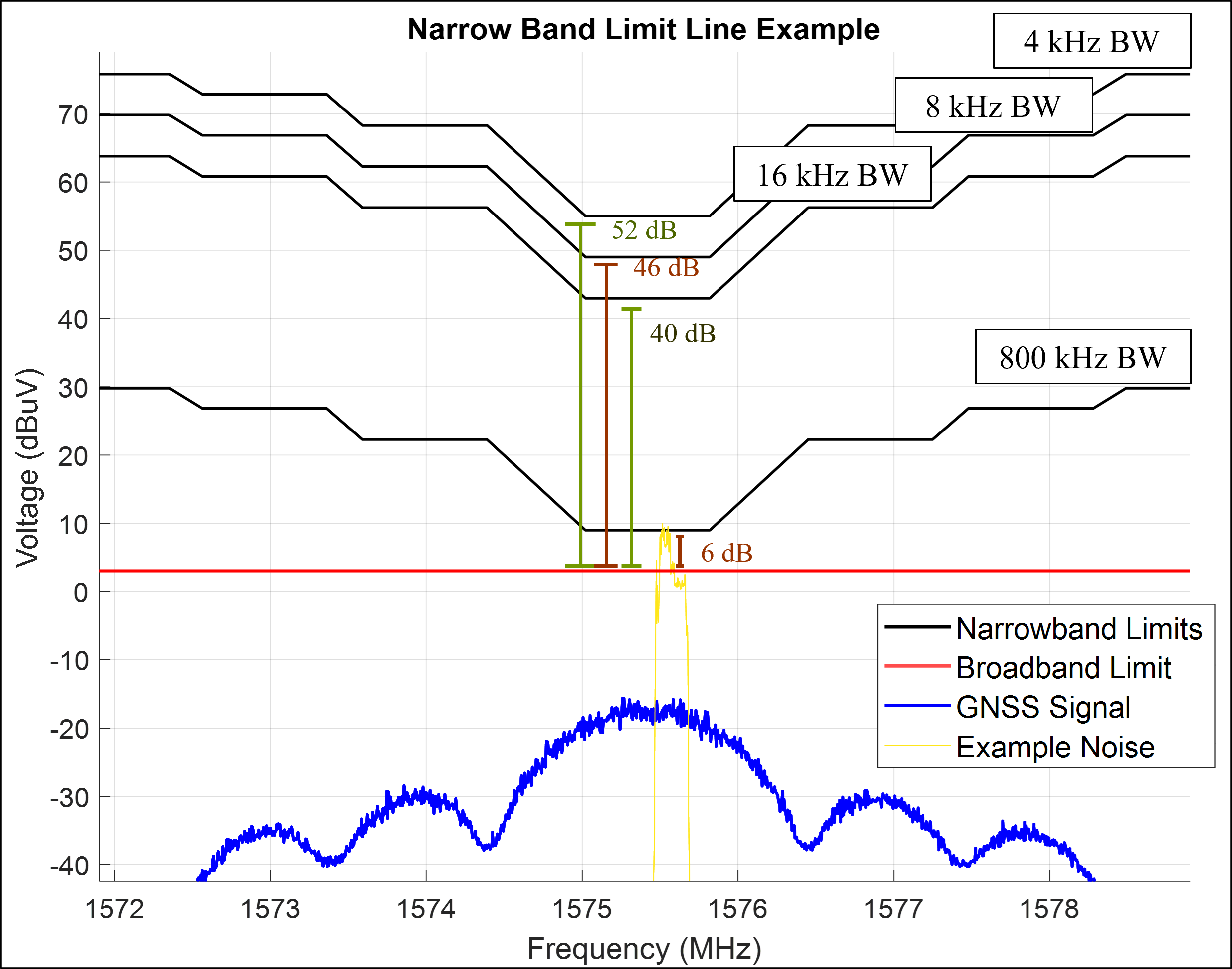}
  \caption{Summary of limitl levels for narrowband and mesoband noise types, broadband noise is chosen such that is drops $C/N_0$ to the point where lock on satellite is lost}
  \label{fig:limit_levels}
\end{figure}

These results indicate that (\ref{eq:mesoband_integrator_output}) can be used as a basis for developing emissions guidelines for mesoband noise sources. Generally, the lowest acceptable value of $C/N_0$ is 35\ dB/Hz. For the measurements in section \ref{Measurement}, the front-end antenna and LNA were bypassed and the signal plus noise was injected directly to the correlator. As a result, the signal powers were roughly 55\ dB higher than the typical GPS L1 signals since the LNA had 55\ dB of gain. Accounting for this bypassed gain, the acceptable values of $C/N_0$, and results above, limit levels for  20\ KHz bandwidth mesoband noise centered at the GPS carrier frequency as shown in Fig. \ref{fig:lmit_level_mesoband_const_center_freq_fig},

it is then followed by limit levels for 40\ kHz and 80\ kHz. As 20\ kHz centered at carrier was measured and modeled it is relatively straightforward to identify power level that drop $C/N_0$ to 35\ dB/Hz, then Eq. (\ref{eq:mesoband_integrator_output}) is used to assess limit levels for following 40\ kHz and 80\ kHz. Next step is to follow sinc envelope to characterize limit levels when center frequency of mesoband noise is varied. In case of moving center frequnecy of the mesoband noise with respect to the carrier, limilt levels can be re-evaluated simply by adjusting for sinc envelope value drop as is demonstrated with Fig. \ref{fig:mesoband_550khz_graph} and Fig. \ref{fig:550Khz_comparison_cn} where esentially evaluating how much sinc envelope value drop compared to its maximum value (usually maximum value will be set as 0\ dB), limit levels can be higher by same amount for given bandwidth noise when limit level was originally developed for noise having same center frequency as GPS L1. By evaluating acceptable limit level for initial baseline case, (in this scenario of 20\ KHz centered at GPS L1) for a particular receiver, user then can follow above mentioned guideline steps to estimate acceptable limit level for another mesoband type interference with different center frequency or bandwidth  and center frequency, in this example Fig. \ref{fig:limit_levels} shows summary, that can be used for mesoband interefence. The  \ref{fig:limit_levels} has been developed for the particular receiver by following steps defined above. Reason limit levels could be different for each receiver is due to $N_0$ ( intrinsic thermal noise), RF front-end noise figure and other receiver specific parameters. The summary of how to construct the limit-levels similar to Fig. \ref{fig:limit_levels} for a particular receiver is as follows: First inject baseline mesoband noise (in our example 20 KHz at center frequency) in addition to actual GPS L1 signal to characterize at what noise power level does the carrier-to-noise ratio drops below acceptable level, note however that at this point engineers also know power level of injected test GPS signal. This can be done using low cost software defined radio and open-source Gitub repository as described in section \ref{Measurement}. Then limil-level similar to Fig. \ref{fig:limit_levels} can be obtained by following sinc envelope. While in EMC testing for systems, if there is a narrowband noise is found to violate CISPR-25 limit leves, engineers can re-evaluate influence of that particular noise to receiver simply by compareing it to baseline case (power level, bandwidth and cener frequnecy) to asses whether system passes or fails EMC testing. However, to be able ot handle arbitrary case, engineers must characterize the GPS receiver in similar fashion by using broadband (usually 50 MHz for GPS L1) rectangular spectal power density  noise and in case of EMC testing, compare arbitrary broadband noise effect on receiver performance by comparing it to rectangular baseline case using eq. \ref{eq:broadband_narrowband_comparison}.

%\IEEEpubidadjcol

\section{Partial and Broadband Interference}
\begin{figure}
  \centering
  \includegraphics[width=0.5\textwidth]{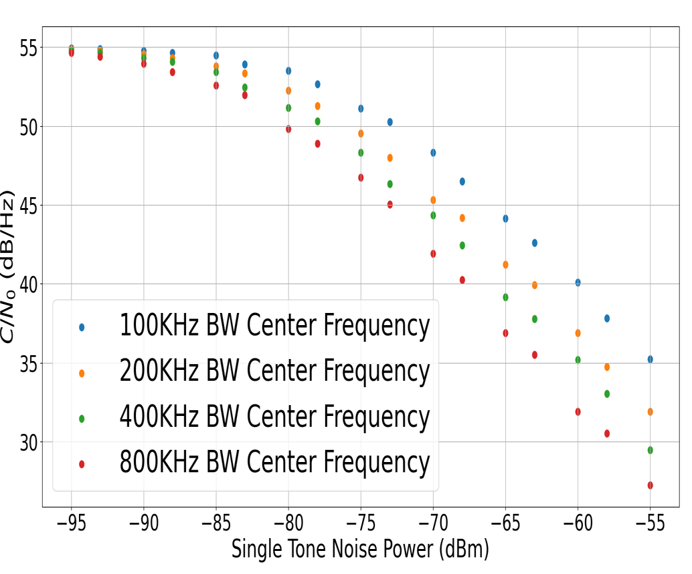}
  \caption{Impact of partial band noise with bandwidth from 100\ kHz to 800\ kHz centered at the carrier frequency of the GPS signal.}
  \label{fig:100_800khz}
\end{figure}

\begin{figure}
  \centering
  \includegraphics[width=0.5\textwidth]{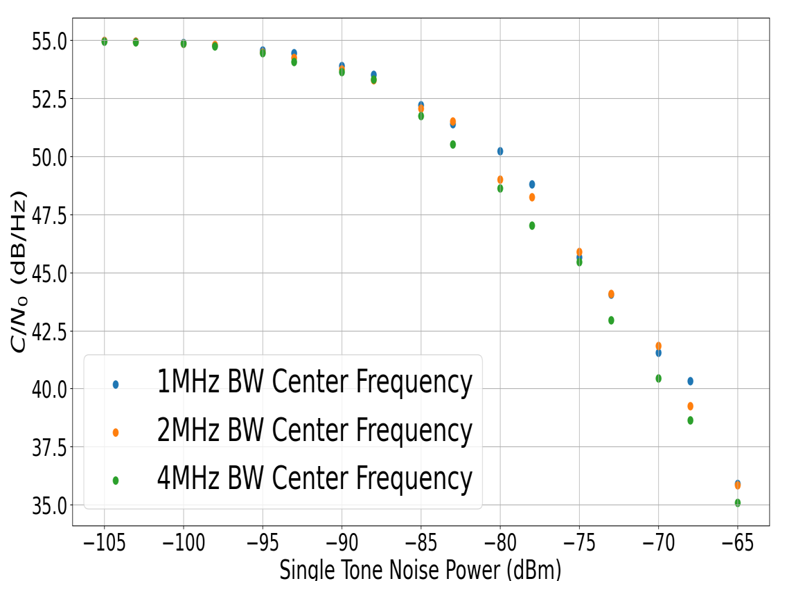}
  \caption{Impact of partial band noise with bandwidth from 1\ MHz to 4\ MHz centered at the carrier frequency of the GPS signal. Beyond 1\ MHz, increasing the bandwidth of the noise has increasing less impact on $C/N_0$. }
  \label{fig:1mhz_4mhz}
\end{figure}

Partial and broadband interference was considered assuming noise with a simple a rectangular power spectrum density and assessing the value of $C/N_0$ using the model described in Section \ref{Theory}. In this case, where the bandwidth is large compared to the main lobe of the C/A code sinc function, the impact of the noise must be evaluated numerically since a closed-form solution for an integral over a portion of a sinc function does not exist. Fig. \ref{fig:100_800khz} shows the value of $C/N_0$ as a function of noise power for different bandwidths of partial-band interference centered at the GPS carrier frequency. When the bandwidth increases from 100\ kHz to 800\ kHz, the value of $C/N_0$ drops by roughly 8\ dB. This result is intuitive since the main lobe is 2\ MHz, so around the GPS center frequency the correlation between the noise and C/A code is still roughly proportion to the bandwidth (i.e. increases by a factor of about 8, or 9 dB, for this change in bandwidth, notice that  10$\log_{10}( \frac{800 Khz}{100 Khz}$) = ~9dB). For bandwidths from 1\ MHz to 4\ MHz, however, the change in $C/N_0$ with bandwidth is much less profound as shown in Fig. \ref{fig:1mhz_4mhz} because the C/A sinc envelope rapidly drops in value beyond a bandwidth of about 1\ MHz. When broadband noise is assumed to have rectangular power spectrum density, the limit level can be evaluated by re-scaling the result from Fig. (\ref{fig:1mhz_4mhz}). As mentioned numerous times throughout the paper, while performing measurements with the GPS receiver, noise was injected after RF front end, esentially  bypassing  the LNA and antenna of the receiver, the scales are subject to adjustment to take LNA gain into consideration. The adjusted limit level for true constant, level broadband interference is shown on Fig. (\ref{fig:limit_levels}), limit line is chosen such that it would highly distrupt receiver and most likely cause it to drop satellite. In case of broad-band interference that does not have rectangular spectrum, it can be averaged over sinc envelope and compared to the true rectangular broadband case. Relatively simple equation can be used to understand degradation from arbitrary spectrum broadband noise as:

\begin{equation}\label{eq:broadband_narrowband_comparison}
10 \log_{10}(\frac{[\int_{- \infty}^{\infty}{\sqrt{P(f)} sinc(\pi f T_{a}) df]}^2 } {[\int_{- \infty}^{\infty}{ {\sqrt{P_0} sinc(\pi f T_{a}) }}df]^2})
\end{equation} 

Where $P_0$ is the limit-level for broadband shown in Fig. (\ref{fig:limit_levels}) in linear scale. For this particular receiver, $P_0$ has been chosen such that it would cause $C/N_0$ tp drop below 35dB/Hz, it can be evaluated either analytically with \ref{eq:cn_multitone} while choosing $P_n$ to be constant (rectanrular power spectrum density for broadband noise) or be found experimentally by injecting noise through receiver and measuring at what level the carrier to noise ratio drops to the point where receiver drops the satellite. The equation essentially captures how much higher arbitrary specturm broadband noise will degrade $C/N_0$ compared to true constant-level broadband noise.

\section{Conclusion}
In conclusion, analytical model has been validated through measurement within ~2 dB using both contrived noise source (arbitrary waveform generator) and more realistic noise sources from live HDMI cable and switch mode power supply. Analytical modelling and measurements were used to was demonstrate that electromagnetic noise with different bandwidth and center frequency can have significantly different effect on carrier-to-noise ratio of the GPS receiver due to C/A codes having sinc envelope in freqyency domain, prompting developemnt of EMC limit-levels based on the characteristics of the noise itself to intelligently asses acceptable radiation from systems that are intended to integrated near GPS receivers. The developed guidelines and example limit-levels esentially allow EMC engineers to have flexible working framework rather than relying on a single limit-level that tries to work for every time of noise whether narrowband or broadband and could be either too restrictive or too lenient.

%\section*{Acknowledgments}
%This work was supported in part by the National Science
% Foundation (NSF) under Grant IIP-1916535

\newpage

% $\backslash${\tt{includegraphics}} command within an optional argument. (You can create your own custom macro containing the $\backslash${\tt{includegraphics}} command to make things
 
\vspace{11pt}
\begin{comment}

\begin{IEEEbiography}[{\includegraphics[width=1in,height=1.25in,clip,keepaspectratio]{spectral_lines_and_noise.png}}]{Author Author}

Use $\backslash${\tt{begin\{IEEEbiography\}}} and then for the 1st argument use $\backslash${\tt{includegraphics}} to declare and link the author photo.
Use the author name as the 3rd argument followed by the biography text.
\end{IEEEbiography}

$author 1 ends$

    This entire block of text
    is a multi-line comment
    and will not be processed
    by LaTeX.

\begin{IEEEbiography}[{\includegraphics[width=1in,height=1.25in,clip,keepaspectratio]{spectral_lines_and_noise.png}}]{Author Placeholder}

Use $\backslash${\tt{begin\{IEEEbiography\}}} and then for the 1st argument use $\backslash${\tt{includegraphics}} to declare and link the author photo.
Use the author name as the 3rd argument followed by the biography text.
\end{IEEEbiography}
\end{comment}

\vspace{11pt}

\vfill

\end{document}